\documentclass[aps,12pt,prb,floats,showpacs,eqsecnum,]{revtex4}
\usepackage{graphicx}
\usepackage{amsfonts}

\usepackage{bm}
\usepackage{amssymb}
\usepackage{amsmath}

\date{\today}
\begin{document}
\title{Sparse random matrices and vibrational spectra \\ of amorphous solids}
\author{Y.\,M.\,Beltukov and D.\,A.\,Parshin}
\affiliation{Saint Petersburg State Polytechnical University,
195251, Saint Petersburg, Russia \\ e-mail: ybeltukov@gmail.com}
\centerline{\null}
\bigskip

\begin{abstract}
\baselineskip=3.5ex

A random matrix approach is used to analyze the vibrational properties of amorphous solids.
We investigated a dynamical matrix $M=AA^T$ with non-negative eigenvalues $\varepsilon=\omega^2$.
The matrix $A$ is an arbitrary  real $N\times N$ \textit{sparse} random matrix with $n$ independent non-zero elements in each row. The average values $\left<A_{ij}\right>=0$ and dispersion $\left<A_{ij}^2\right>=V^2$ for all non-zero elements. The density of vibrational states $g(\omega)$ of the matrix $M$ for $N,n\gg 1$ is given by the Wigner quarter circle law, $g(\omega)=(N/\pi nV^2)\sqrt{4nV^2-\omega^2}$, with radius independent of $N$.
We argue that for $n^2\ll N$ this model can be used to describe the interaction of atoms in amorphous solids. The level statistics of matrix $M$ is well described by the Wigner surmise and corresponds to repulsion of
eigenfrequencies. The participation ratio for the major part of vibrational modes in three dimensional system is about $0.2 - 0.3$ and independent of $N$. Together with term repulsion it indicates clearly to the delocalization of vibrational excitations. We show that these vibrations spread in space by means of diffusion. In this respect they are similar to diffusons introduced by Allen, Feldman, et al., Phil. Mag. B 79, 1715 (1999) in amorphous silicon. Our results are in a qualitative and sometimes in a quantitative agreement with molecular dynamic simulations of real and model glasses.
\end{abstract}
\maketitle
\baselineskip=3.5ex

\newpage

\section{Introduction}

One of the main problem in the physics of disordered systems is establishing the general properties of vibrations in amorphous solids (glasses). At low frequencies $\omega$ there are usual plane wave excitations (longitudinal and transverse phonons) characterized by well defined wave vector  $\mathbf{k}$. These are delocalized excitations. Their density of states (in three dimensional space) $g_{\rm ph}(\omega) \propto \omega^2$.

According to recent investigations besides phonons in glasses exist low frequency quasilocal vibrational modes \cite{GPS}. Their density of states, $g_{\rm qlv}(\omega)$, is a universal function of $\omega$.
At low frequencies, below some characteristic frequency $\omega_b$, the density of states
$g_{\rm qlv}(\omega) \propto \omega^4$, and at higher frequencies for $\omega > \omega_b$,
$g_{\rm qlv}(\omega) \propto \omega$. As a result the reduced density of states
$g_{\rm qlv}(\omega)/\omega^2$ has a broad maximum at frequency $\omega \simeq \omega_b$. This is the boson peak which was found almost in all glasses. Its theory was developed recently in \cite{GPS,PGS}.

However the range of the boson peak in glasses normally occupies only 10\% of vibrational spectrum. At higher frequencies, for example in amorphous SiO$_2$, the vibrational density of states $g(\omega)$ approaches a constant value which extends towards maximum frequencies (of the order of the Debye frequency $\omega_{D}$)~\cite{Wei}.
The same behavior of $g(\omega)$ was found in the soft-sphere glass~\cite{Schober1}, and in amorphous
Se~\cite{Schober93}. In other glasses the density of states has a broad maximum around $\omega_D$ and then decays to zero~\cite{Hafner,Meshkov,Ballone,Abraham}.

The nature of these higher frequency excitations which dominate in the vibrational spectra is not yet known and has been discussed a lot in the literature~\cite{Diffusons,Nature}. The main question is whether these vibrations are plane waves (phonons) or not? And if they are not phonons, whether they are localized or delocalized? The solution of this crucial question is very important, for example, for elucidating the nature of the thermal conductivity of amorphous dielectrics in this frequency (or temperature) range.

For example the numerical studies  of amorphous silicon~\cite{Nature} show that the lowest
4\% of vibrational modes are plane-wave like ({\em propagons}, or phonons) and the highest 3\% of modes are localized ({\em locons}). The rest are neither plane-wave like nor localized. They were called {\em diffusons}, having in mind  that they spread in space by means of diffusion (from one atom to another)~\cite{Nature}.
Many computer simulations of real and model glasses show that the most part of vibrational modes are indeed delocalized~\cite{Wei,Schober1,Schober93,Hafner,Meshkov,Ballone,Abraham}.

One of the goals of this work is to answer the crucial question what is the nature of vibrational excitations in disordered systems and in particular of diffusons using the methods of random matrix theory. These methods were very successful in understanding of universal properties of electronic disordered systems, in theory of financial markets, random networks, wireless communications, etc. However there are very small amount of papers using these methods for investigation of vibrations in disordered systems (see for example~\cite{Chalker}).

We would like to fill this gap. However we do not want to attach ourselves to some specific disordered system (amorphous solid, liquid or glass). We will consider the dynamical vibrational matrix of a general form which is a symmetric random matrix with positive eigenvalues.  Therefore we hope that properties of this matrix will be sufficiently general and independent of particular details of vibrational system. Some of these properties may resemble those in real glasses.

In the present paper we restrict ourselves to vibrational disordered systems where are no delocalized low frequency plane wave excitations (phonons). Therefore if vibrational modes in our system will happen to be delocalized it would have no relation with phonons. In fact for existence of phonons it should exist some order in the disordered system. Therefore, we consider this case  separately somewhere~\cite{Beltukov1}.

\section{Random matrix approach to vibrations in disordered systems}
\label{rma}

Newton equations of mechanical vibrational system in harmonic approximation can be presented in a form (scalar  model)~\cite{Maradydin}
\begin{equation}\label{dynamic}
     m_i\omega^2u_i=\sum\limits_{j=1}^N\Phi_{ij}u_j ,\quad i=1,2,...,N .
\end{equation}
Here $m_i$ and $u_i$ are masses and atomic displacements correspondingly, $\Phi_{ij}$ is a force constant matrix (Hessian). It is a real, symmetric and positive definite matrix. Introducing the new variables
\begin{equation}
     q_i=u_i\sqrt{m_i} ,\quad M_{ij}=\Phi_{ij}/\sqrt{m_im_j}
     \label{new}
\end{equation}
we come to the equations
\begin{equation}
     \omega^2q_i=\sum\limits_{j=1}^N M_{ij}q_j ,\quad i=1,2,...,N. \label{dynamic3}
\end{equation}
The dynamical matrix $M$, as well as matrix $\Phi$ is a real, symmetric and positive definite matrix.
At some values $\varepsilon = \omega^2$ which are eigenvalues of the dynamical matrix, the system (\ref{dynamic3}) has non-zero solutions. These values of $\omega$ are vibrational eigenfrequencies of the mechanical system.

To describe vibrations in amorphous solids we are going to apply methods of random matrix theory. At first glance it is a rather difficult mathematical problem.  Not every random symmetric dynamical matrix is appropriate for studying
vibrations. Matrix $M$ should be also {\em positive definite}. It ensures mechanical stability of the system.
For that, matrix elements $M_{ij}$ should be correlated with each other in a non trivial way. Therefore they cannot be taken as independent random numbers. We will solve this problem using a following mathematical approach.

Every real symmetric and positive definite matrix $M$ one can always present in the form~\cite{PositiveDefine}.
\begin{equation}
     M = AA^T.
\end{equation}
Here $A$ is some real (not necessary symmetric) matrix of a general form. And vice versa for every real matrix $A$
the product $AA^T$ is a positive definite symmetric matrix~\cite{PositiveDefine}. One may assume that in amorphous solids (in some frequency interval) $\varepsilon = \omega^2$ are eigenvalues of matrix $AA^T$, where $A$ is taken in some way randomly.

Distribution of eigenvalues for matrix $M=AA^T$ for some classes of random matrices $A$ was first derived in the
paper of Marchenko and Pastur~\cite{Marchenko}. Then it was investigated in the theory of financial markets~\cite{Plerov}, complex networks~\cite{Barthelemy} and wireless communications~\cite{Tulino}. As far as we know for vibrations in disordered solids this approach was not used so far (as an exception see paper~\cite{Chalker}).
In the papers~\cite{Plerov,Barthelemy,Tulino} it was mainly investigated the case of Wishart ensemble~\cite{Wishart} when matrix $A$ is a random matrix with independent matrix elements with zero mean $\left\langle A_{ij}\right\rangle = 0$ and equal dispersion $\left\langle A_{ij}^2\right\rangle = V^2$.

It was shown in these papers that for a real matrix $A$ (dimensions $N\times N$) the distribution of frequencies $\omega$ (square roots from eigenvalues $\varepsilon$ of the matrix $M=AA^T$) for $N\gg 1$ has a {\em quarter circle}
form
\begin{equation}
     g(\omega) = \frac{1}{\pi V^2}\sqrt{4NV^2-\omega^2}, \quad 0 < \omega <  2V\sqrt{N} .
     \label{solve32}
\end{equation}
It formally coincides with the well known Wigner semicircle~\cite{Wigner}
for distribution $g(\varepsilon)$ of the eigenvalues $\varepsilon$
of symmetric random matrix $H$ with independent random matrix elements with zero mean $\left\langle H_{ij}\right\rangle = 0$ and equal dispersion $\left\langle H_{ij}^2\right\rangle = V^2$. In this case Eq.~(\ref{solve32}) with additional coefficient $1/2$ and replacement $\omega$ to  $\varepsilon$ is valid in the interval
$-2V\sqrt{N} < \varepsilon < 2V\sqrt{N}$.

In the random medium model described by the Wishart ensemble, each of the elements $M_{ij}$ of the dynamical matrix
$M$ is not zero
\begin{equation}
M_{ij}=\sum_k A_{ik} A_{jk}.
\label{vst5rf}
\end{equation}
Obviously this corresponds to the case of {\em long-range} interaction, when each atom is connected by random springs with every atom in the system. However, this model is not justified from the physical
point of view. In amorphous materials, only closely
spaced atoms are bound together by elastic springs. Therefore the more real would be the case when the number of nonzero elements $m$ in each row (and column) of the matrix $M$ is small as compared to $N$ and does not depend on $N$. As a result, the matrix $M$ should be sparse. Actually, such sparse matrices arise in computer calculations of atomic vibrations in
amorphous solids (and liquids). For example, in the case of the short-range order for a simple cubic lattice with the
interaction only between nearest neighbors and the vector character of vibrations (in three dimensional
space), we have $m = 18 + 3 = 21$. For other lattices we have $m = 24 + 18 + 3 = 45$ for a bcc lattice and $m = 36 + 18 + 3 = 57$ for a fcc  lattice. In the last two cases we took into account all interactions in the first and second coordination shells.

Therefore we get a more real case if we consider a sparse matrix $A$ where each row contains only $n$ nonzero matrix elements (with $n\ll N$). Then, each row of the matrix $M=AA^T$ will have, on average, $m = n^2$ nonzero elements. At $n^2\ll  N$, this corresponds to the case of a sparse matrix $M$. We will show below, that in this case, for $n\gg  1$, the density of states is also described by the quarter-circle law
\begin{equation}
     g(\omega) = \frac{N}{\pi nV^2}\sqrt{4nV^2-\omega^2}, \quad 0 < \omega <  2V\sqrt{n}
     \label{duvb6}
\end{equation}
but with radius independent of $N$. The derivation of formulas (\ref{solve32}) and (\ref{duvb6}) is given below,
because in this form (and as applied to the problem of vibrations) it is absent in the literature. The derivation
given in~\cite{Marchenko} is mathematically rather difficult for non-specialists.

\section{The quarter-circle law}
\label{par:quarter}

First let us derive Eq.~(\ref{solve32}). We assume that $A$ is a random square matrix
$N\times N$ (with $N\gg 1$) of a
general form (not necessarily symmetric) with independent random matrix elements $A_{ij}$ so that
for any $i$ and $j$.
\begin{equation}
     \langle A_{ij} \rangle = 0, \quad \left\langle { A_{ij} }^2 \right\rangle = V^2.
     \label{ser3}
\end{equation}
To derive the distribution density of eigenvalues for matrix $M = AA^T$ we will use the method similar to the derivation of the distribution density of eigenvalues of a random symmetric matrix~\cite{Mehta}.

For that we fix matrix $A$ and add to it an independent small random matrix $\delta A$ with all its
elements being independent so that for any $i$ and $j$
\begin{equation}
     \langle \delta A_{ij} \rangle = 0, \quad \left\langle  \delta A_{ij} ^2   \right\rangle = v^2,
\end{equation}
with $v \ll V$. Then the matrix $M$ is changed by
\begin{equation}
     \delta M = (A+\delta A)\cdot(A+\delta A)^T-M=
      \delta A \cdot A^T + A \cdot \delta A^T + \delta A \cdot \delta A^T.
     \label{deltaM1}
\end{equation}
Let $U$ be an orthogonal matrix that transforms the matrix $M$ into a diagonal form. In this representation the new matrix $\delta \widetilde M$ can be defined as
\begin{equation}
     \delta\widetilde{M}=U^{-1} \cdot \delta M \cdot U.
     \label{UTMU1}
\end{equation}
According to the perturbation theory the change of the $i$th eigenvalue $\varepsilon_i$ of the matrix $M$ (due to the addition of the matrix $\delta M$) can be written in the form
\begin{equation}
     \delta\varepsilon\left(\varepsilon_i, V^2\right)=\delta\widetilde{M}_{ii} + \sum_{j\neq i}{\frac{{\delta\widetilde{M}_{ij}}^2}{\varepsilon_i-\varepsilon_j}}+\cdots.
     \label{Perturbation2_1}
\end{equation}

Let us average this equation over random matrix $\delta A$. The matrix $A$ we keep fixed therefore $U$, $M$, and eigenvalues $\varepsilon_i$ are also fixed. For this purpose we have to calculate the mean value and the variance of the elements of matrix $\delta \widetilde{M}$. The mean value of the diagonal elements is given by
\begin{equation}
     \left\langle\delta\widetilde{M}_{ii}\right\rangle = \Big\langle\sum_{klm}U_{ki}U_{li}\big(A_{km}\delta A_{lm}+
     \delta A_{km} A_{lm}+\delta A_{km}\delta A_{lm}\big)\Big\rangle.
     \label{deltaMiifirst}
\end{equation}
The first and second terms in parentheses are equal on average to zero since $\left<\delta A \right>=0$.
The third term at $k\neq l$ is also equal on average to zero but at $k = l$ it is equal to $v^2$.
As a result we get
\begin{equation}
     \left\langle\delta\widetilde{M}_{ii}\right\rangle=\sum_{km} U_{ki} ^2\left\langle{\delta A_{km}}^2\right\rangle=v^2\sum_{km}{U_{ki}}^2=v^2N.
     \label{deltaMii}
\end{equation}
In the last equality we used orthogonality of the matrix $U$.

The variance of the off-diagonal elements $(i \neq j)$ is given by
\begin{equation}
     \left\langle {\delta\widetilde{M}_{ij}}^{2}\right\rangle = \Big\langle\Big(\sum_{klm}U_{ki}U_{lj}(A_{km}\delta A_{lm}+
     \delta A_{km}A_{lm}+\delta A_{km}\delta A_{lm})\Big)^2\Big\rangle.\label{deltaMijSqr1}
\end{equation}
We can neglect the last term in the parentheses because it leads to the higher order corrections. Then
\begin{eqnarray}
     \left\langle {\delta\widetilde{M}_{ij}}^{2}\right\rangle =  \Big\langle\Big(\sum_{klm}U_{ki}U_{lj}A_{km}\delta A_{lm}\Big)^2\Big\rangle+
     \Big\langle\Big(\sum_{klm}U_{ki}U_{lj}\delta  A_{km}A_{lm}\Big)^2\Big\rangle +\nonumber\\
     +2\Big\langle\sum_{klm}U_{ki}U_{lj}A_{km}\delta A_{lm} \cdot \sum_{klm}U_{ki}U_{lj}\delta A_{km}A_{lm}\Big\rangle.
     \label{deltaMijSqr2}
\end{eqnarray}
In the first term we expand the square. The nonzero contribution is given only by the terms that contain
the product of the same elements of the matrix $\delta A$. Their mean is equal to $v^2$
\begin{equation}
     \Big\langle\Big(\sum_{klm}U_{ki}U_{lj}A_{km}\delta A_{lm}\Big)^2\Big\rangle   =
      \sum_{k_1 k_2 l m}U_{k_1 i}U_{k_2 i}{U_{lj}}^2 A_{k_1 m} A_{k_2 m} v^2.
     \label{deltaMijSqrTerm1first}
\end{equation}
The sum over $l$ of ${U_{lj}}^2$ gives $1$ due to orthogonality conditions for matrix $U$.
The reminder is the diagonal element ($ii$) of the matrix $U^{-1}AA^TU$ which is the eigenvalue $\varepsilon_i$.
Therefore the first term in Eq.~(\ref{deltaMijSqr2}) is equal to $v^2\varepsilon_i$.
By renaming the indices we similarly find that the second term in Eq.~(\ref{deltaMijSqr2}) is $v^2\varepsilon_j$.
Because of orthogonality of the matrix $U$, the third term in Eq.~(\ref{deltaMijSqr2}) is equal to zero. As a result we get
\begin{equation}
     \left\langle {\delta\widetilde{M}_{ij}}^{2}\right\rangle =  v^2(\varepsilon_i+\varepsilon_j).
     \label{deltaMijSqr3}
\end{equation}

Let us now insert Eqs.~(\ref{deltaMii}, \ref{deltaMijSqr3}) into the right-hand side of averaged Eq.~(\ref{Perturbation2_1}). As a result the mean value of the correction to the $i$th eigenvalue can be  written in the form (we will keep the same notation for it)
\begin{equation}
     \delta\varepsilon\left(\varepsilon_i, V^2\right)=v^2N+v^2\sum_{j \neq  i}\frac{\varepsilon_i+\varepsilon_j}{\varepsilon_i-\varepsilon_j}.
     \label{Perturbation3}
\end{equation}
Taking into account that $N\gg 1$, we include the first term into the sum
\begin{equation}
     \delta\varepsilon\left(\varepsilon_i, V^2\right)=2v^2\varepsilon_i\sum_{j \neq i}\frac{1}{\varepsilon_i-\varepsilon_j}. \label{Perturbation4}
\end{equation}
For $N\gg 1$, the distribution density $\sigma(\varepsilon, V^2)$ of eigenvalues of the matrix $M$ is a continuous
function of $\varepsilon$. Then sum (\ref{Perturbation4}) can be approximated by the principal value of the integral
\begin{equation}
     \delta\varepsilon\left(\varepsilon, V^2\right)=2v^2\varepsilon\int\frac{\sigma(\varepsilon', V^2)}{\varepsilon-\varepsilon'}\,d\varepsilon'. \label{Perturbation5}
\end{equation}

Now let us consider a change of the number of eigenvalues of the matrix $M$ in the interval $(\varepsilon,
\varepsilon+\Delta\varepsilon)$ due to the change of the matrix $A$ to the matrix $A+\delta A$.
On the one hand this change is given by
\begin{equation}
     \sigma\left(\varepsilon, V^2\right)\delta \varepsilon\left(\varepsilon,V^2\right)-
     \sigma\left(\varepsilon+\Delta\varepsilon, V^2\right)\delta \varepsilon\left(\varepsilon+\Delta\varepsilon, V^2\right)\approx
      -\frac{\partial(\sigma \delta\varepsilon)}{\partial\varepsilon}\Delta \varepsilon.
     \label{inc1}
\end{equation}
On the other hand since we added an independent random matrix $\delta A$ to the matrix $A$, the variance of the
elements of the new matrix $A+\delta A$ increased from $V^2$ to $V^2+v^2$. Therefore the change of the number of eigenvalues in the interval $(\varepsilon, \varepsilon+\Delta\varepsilon)$ can be presented in the form
\begin{equation}
     \sigma\left(\varepsilon, V^2+v^2\right)\Delta\varepsilon-\sigma\left(\varepsilon, V^2\right)\Delta\varepsilon \approx v^2\frac{\partial\sigma}{\partial(V^2)}\Delta\varepsilon.
     \label{inc2}
\end{equation}
By equating (\ref{inc1}) and  (\ref{inc2}) we obtain the second equation
\begin{equation}
     -\frac{\partial(\sigma \delta\varepsilon)}{\partial\varepsilon} = v^2\frac{\partial\sigma}{\partial(V^2)}. \label{incinc}
\end{equation}

To solve the system of two equations (\ref{Perturbation5}) and (\ref{incinc}) let us introduce the new variable $x=\varepsilon/V^2$.
Since eigenvalues of the matrix $M$ are proportional to $V^2$, the quantity $x=\varepsilon/V^2$ for each eigenvalue
remains unchanged when all elements of the matrix $A$ are multiplied by a constant. Secondly, according to the normalization condition, the density of eigenvalues is scaled as $1/V^2$. As a result the distribution density can be presented in the form
\begin{equation}
     \sigma\left(\varepsilon, V^2\right) = \frac{1}{V^2}\,\widetilde{\sigma}(x),
     \label{tildeSigma}
\end{equation}
where $\widetilde{\sigma}(x)$ is some function of $x$. Making use of it we find from Eq.~(\ref{Perturbation5}) that $\delta \varepsilon$ can be presented in the form
\begin{equation}
     \delta \varepsilon\left(\varepsilon, V^2\right) = v^2\widetilde{\delta
       \varepsilon}(x).\label{tildeDeltaEpsilon}
\end{equation}
Then Eq.~(\ref{Perturbation5}) transforms as follows
\begin{equation}
     \widetilde{\delta\varepsilon}(x) = 2x\int\frac{\widetilde{\sigma}(x')}{x-x'}dx'. \label{intEq}
\end{equation}

Substitution of Eqs.~(\ref{tildeSigma}, \ref{tildeDeltaEpsilon})  into Eq.~(\ref{incinc}) gives us the differential equation with one variable
\begin{equation}
     \frac{\partial \left(\widetilde{\sigma}\widetilde{\delta\varepsilon}\right)}{\partial x} = \frac{\partial \big(x\widetilde{\sigma}\big)}{\partial x}. \label{diffEq}
\end{equation}
It follows from Eq.~(\ref{intEq}) that $\widetilde{\delta\varepsilon}(0) = 0$. As a result after integration of Eq.~(\ref{diffEq}) we get
\begin{equation}
     \widetilde{\delta\varepsilon}(x) = x. \label{diffEqEasy}
\end{equation}
Finally using that  we derive from Eq.~(\ref{intEq}) the integral equation
\begin{equation}
     \int\frac{\widetilde{\sigma}(x')}{x-x'}dx'=\frac{1}{2}.
     \label{intEq2}
\end{equation}
The normalization condition can be written in the form
\begin{equation}
     \int\sigma\left(\varepsilon, V^2\right)d\varepsilon=\int\widetilde{\sigma}(x)dx=N.
     \label{norm}
\end{equation}

The solution of Eq.~(\ref{intEq2}) with the normalization condition (\ref{norm}) is given by
\begin{equation}
     \widetilde{\sigma}(x) = \frac{1}{2\pi}\sqrt{\frac{4N-x}{x}}, \quad 0<x<4N.
     \label{solve}
\end{equation}
Coming back to the original variables we get the distribution density of eigenvalues for the matrix $M$
\begin{equation}
     \sigma\left(\varepsilon, V^2\right) = \frac{1}{2\pi
       V^2}\sqrt{\frac{4NV^2-\varepsilon}{\varepsilon}}, \quad 0<\varepsilon<4NV^2.
     \label{solve2}
\end{equation}
This distribution is a particular case of the Marchenko-Pastur distribution~\cite{Tulino,Marchenko}.
Taking into account that $\varepsilon = \omega^2$ the density of vibrational states can be represented as
\begin{equation}
     g(\omega) = \frac{1}{\pi V^2}\sqrt{4NV^2-\omega^2}, \quad 0 < \omega < 2V\sqrt{N}. \label{solve3}
\end{equation}
This equation has the form of a quarter-circle.

As was already mentioned above in Section~\ref{rma} this model with the long-range interaction is not justified from the physical point of view for vibrations in amorphous solids. On the one hand it implies  that each atom interacts with every atom in the system. On the other hand it gives the width of the vibrational spectrum depending on the size of the system $N$. In the next section we consider the sparse random matrix $A$. In such a way one can overcome these two limitations.

\section{A sparse random matrix}

Let us consider a sparse random matrix $A$. Now in each row $i$ we randomly choose $n$ positions $j$ and
write random numbers $a_{ij}$ there with the same probability density $\rho_0(a_{ij})$. In all other positions we write zeros. As before all nonzero random matrix elements are characterized by their mean $\left\langle a_{ij}\right\rangle = 0$ and the variance $\left\langle a_{ij}^2\right\rangle = V^2$. An example of such sparse random matrix $A$ ($8\times  8$) with $n = 3$ is shown below
\begin{equation}
     A = \left(
          \begin{array}{cccccccccc}
               0 & 0 & * & 0 & * & 0 & 0 & *\\
               \vphantom{x} * & * & 0 & 0 & 0 & 0 & 0 & *\\
               0 & 0 & * & 0 & * & 0 & * & 0\\
               \cdots & \cdots & \cdots & \cdots & \cdots & \cdots & \cdots & \cdots\\
               0 & * & 0 & 0 & * & * & 0 & 0
          \end{array} \right).
\end{equation}
The asterisks here indicate nonzero random matrix elements.

The elements of such random matrix are independent from each other and their distribution is given by the probability density
\begin{equation}
     \rho(x) = \frac{n}{N}\rho_0(x)+\frac{N-n}{N}\delta(x).
     \label{uu78}
\end{equation}
Here $\delta(x)$ is the Dirac delta function. The mean value of matrix elements $\left<A_{ij}\right>=0$. The variance of this distribution is given by
\begin{equation}
     \left\langle A_{ij}^2\right\rangle = \int x^2\rho(x)\,dx = \frac{n}{N}\int x^2\rho_0(x)\,dx = \frac{n}{N}V^2.
\end{equation}
The averaging is performed over both zero and nonzero elements of the matrix $A$.

Such sparse random matrix $A$ satisfies to conditions similar to Eqs.~(\ref{ser3}) used for derivation of Eq.~(\ref{solve3}). Therefore, for $n\gg 1$ we can use the derived quarter-circle density of states (\ref{solve3}) by substituting the variance $nV^2/N$ instead of the variance $V^2$. As a result
\begin{equation}
     g(\omega) = \frac{N}{\pi n V^2}\sqrt{\big.4nV^2-\omega^2}, \quad 0< \omega <
     2V\sqrt{n}.
     \label{xc6}
\end{equation}
At the same time we can take $n\ll N$. This means that for $N\gg n\gg 1$ the quarter-circle distribution for the density of states $g(\omega)$ is still valid even in the case when nonzero elements occupy only a small part of the matrix $A$. Such form of the distribution $g(\omega)$ in our model is a universal law and does not depend on the distribution density $\rho_0(a_{ij})$, the size of the system $N$, and the number of nonzero elements $n$ for sufficiently large values of $n$.

\begin{figure}[!here]
     \centering
     \includegraphics[width=0.6\textwidth]{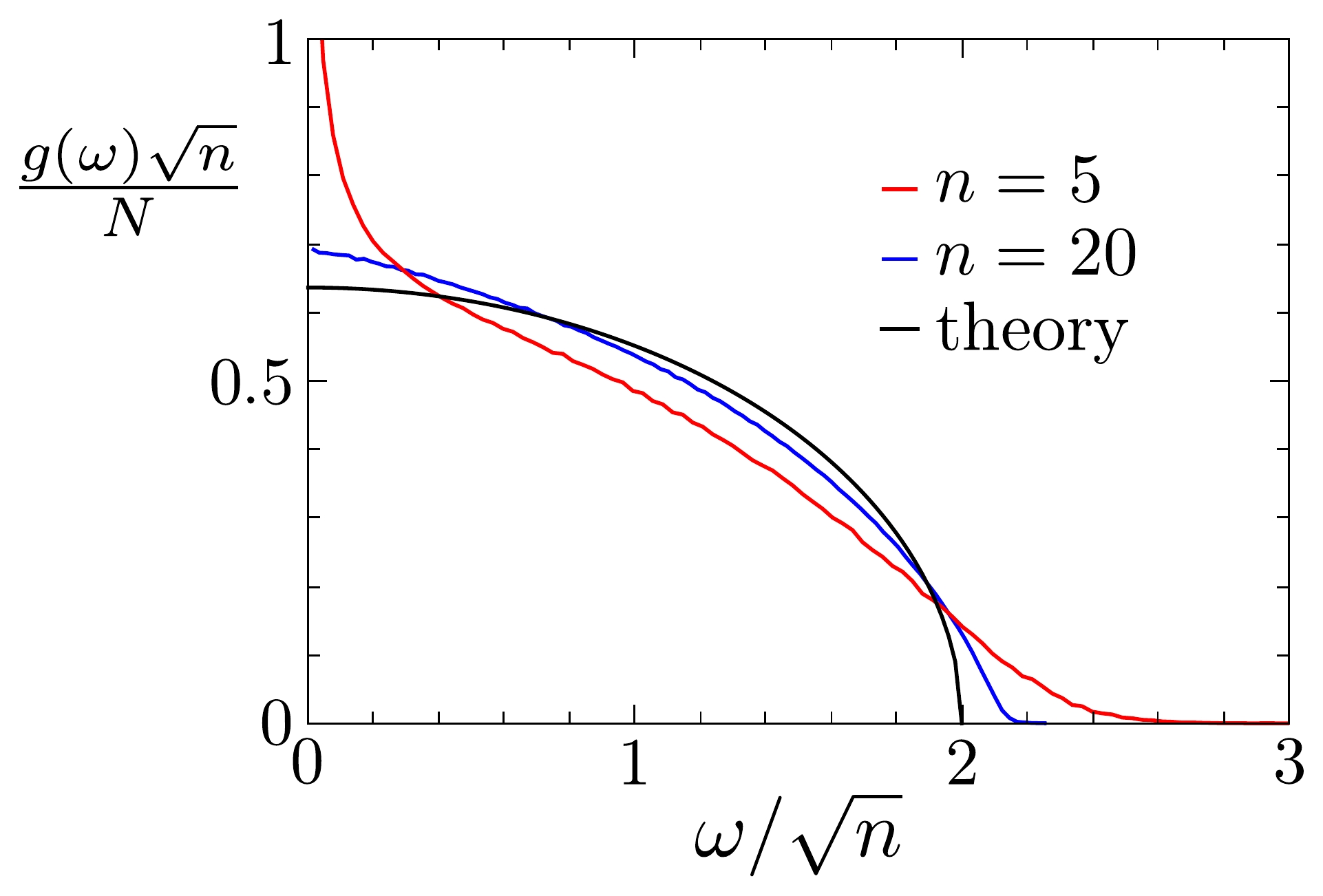}
     \caption{Numerically calculated density of states $g(\omega)$ of the matrix $M = AA^T$ for two different values of $n$ and $N=1000$.
      Black line shows the density of states given by Eq.~(\ref{xc6}). The random matrix $A$ has a normal Gaussian distribution of nonzero matrix elements $\rho_0(a_{ij})$ with $\left\langle a_{ij}\right\rangle = 0$ and variance         $\left\langle a_{ij}^2\right\rangle = 1$.}
       \label{fig:some_a}
\end{figure}
The  numerical analysis confirms that with an increase of $n$  the density of
states $g(\omega)$ indeed approaches the quarter-circle distribution (for $n \gg 1$) and in this case the inequality $n \ll N$ is possible (see Fig.~\ref{fig:some_a}). Already for $n$ of the order of $10$ and $N=1000$, we get the density of states that is only slightly different from the quarter-circle distribution.
Normalized to unity it does not depend on the size of the system $N$.

For such sparse random matrix $A$ with $n$ nonzero elements in each row, each row (and each column) of the symmetric matrix $M$ has on average $n^2$ nonzero elements (for $n^2\ll N$).
Physically this implies that one atom interacts with approximately $n^2$ close atoms. The possibility to take a fixed value of $n\gg 1$ and allowing $N \to \infty$ gives us the opportunity to describe the interaction between close atoms in amorphous materials what is physically reasonable.

From Eq.~(\ref{xc6}) we can obtain the distribution $p(\varepsilon)$ of the eigenvalues $\varepsilon=\omega^2$ of the matrix $M=AA^T$
\begin{equation}
\label{vc66}
p(\varepsilon)=\frac{N}{2\pi n V^2}\sqrt{\frac{4nV^2-\varepsilon}{\varepsilon}} .
\end{equation}
This expression has a singularity at $\varepsilon\to 0$ ($p(\varepsilon)\propto 1/\sqrt{\varepsilon}$).
Probably this singularity in the distribution of eigenvalues of the dynamical matrix was observed in~\cite{Taraskin} for the amorphous and liquid phases of $\rm{SiO_2}$. But it was attributed (improperly in our opinion) to special correlations between diagonal and non-diagonal elements of the dynamical matrix. A similar behavior at $\varepsilon\to 0$ was observed for the distribution of eigenvalues in~\cite{Huang} (see Fig.~2 of this paper).

In conclusion of this section, it should be noted that the singularity in the density of states $g(\omega)$ at
$\omega\to 0$ manifests itself for small values of $n$ (see Fig.~\ref{fig:some_a} for n = 5).
Similar singularity exists also in the density of eigenvalues $g(\varepsilon)$ (at $\varepsilon\to 0$) of the symmetric sparse random matrix $H_{ij}$ for small values of $n$. The singularity is suppressed with increase of $n$ when the density of eigenvalues $g(\varepsilon)$ approaches to the form of  Wigner semicircle~\cite{Dyson,Rodgers,Evangelou_1,Evangelou_2}. This singularity was
first discovered in the density of vibrational states of a disordered one-dimensional chain by Dyson~\cite{Dyson}. Therefore it sometimes is called the Dyson singularity. It has been believed that this singularity is an indication of strong fluctuations in a random medium and related localization of modes~\cite{Evangelou_2}. We hope to consider this problem in more details in a separate work.

\section{The cubic lattice with random bonds}
\label{randcub}

The symmetric sparse random matrix $M = AA^T$ considered in the previous section is topologically
equivalent to a {\em tree} (closed to itself on the system size). The number $m = n^2$ specifies the order of
branching or the coordination number of this tree. However a random bonds structure in amorphous
solids (glasses) more likely corresponds to the short-range order in the atomic arrangement topologically
similar to the bond structure existing in the corresponding crystals. It is clear that topologically a crystal structure differs fundamentally from a tree structure. In a tree structure there are no closed loops of bonds which are present in a lattice. Therefore our purpose in this section is to generalize obtained results to spatial structures with the topology of a crystal lattice rather than a tree. Otherwise our system remains random without any periodicity (except for the topology of the elastic bonds).

Let us consider how making use of a random matrix $A$ we can build a dynamical random matrix $M=AA^T$ with the known topological bond structure but having random bond strengths. As an example we consider a simple cubic lattice
$w\times w\times w$ with $N=w^3$ atoms. The atoms have integer coordinates $(x, y, z)$ and each coordinate can take on values from $1$ to $w$. Let us introduce the integer index $i=x+w(y-1)+w^2(z-1)$. Each atom in the lattice is characterized by its unique index $i$ running from $1$ to $N$.

\begin{figure}[here!]
     \centering
     \includegraphics[width=0.50\textwidth]{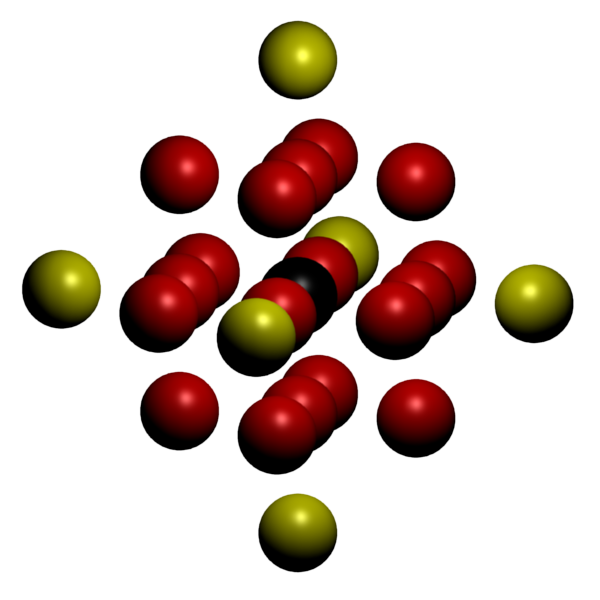}
    \caption{Schematic diagram illustrating the interaction of atoms in a simple cubic lattice. Shown are atoms interacting with the central (black) atom. Yellow atoms have on average a less rigid bonds with the central atom.}
     \label{fig:Atoms}
\end{figure}

We construct a random matrix $A$ in the following way. The element $A_{ij}$ is nonzero and random if the $i$th and $j$th atoms are nearest neighbors or it is the same atom (for $i=j$). All other elements $A_{ij}$ we take equal to zero. We note that for $i\ne j$ the elements $A_{ij}$ and $A_{ji}$ are independent random numbers (the matrix $A$ is not symmetric). As a result for a simple cubic lattice we have seven nonzero elements in each row and in each column of matrix $A$ (except for the rows and columns corresponding to boundary atoms).
In the dynamical matrix $M=AA^T$ the element $M_{ij}$ will be nonzero if $i$th atom will be nearest or next nearest to the $j$th atom (or it is the same atom for $i=j$). Fig.~\ref{fig:Atoms} shows the atoms interacting with the central (black) atom. All other atoms in the lattice interact similarly with their neighbors.

\begin{figure}[!here]
     \centering
     \includegraphics[width=0.50\textwidth]{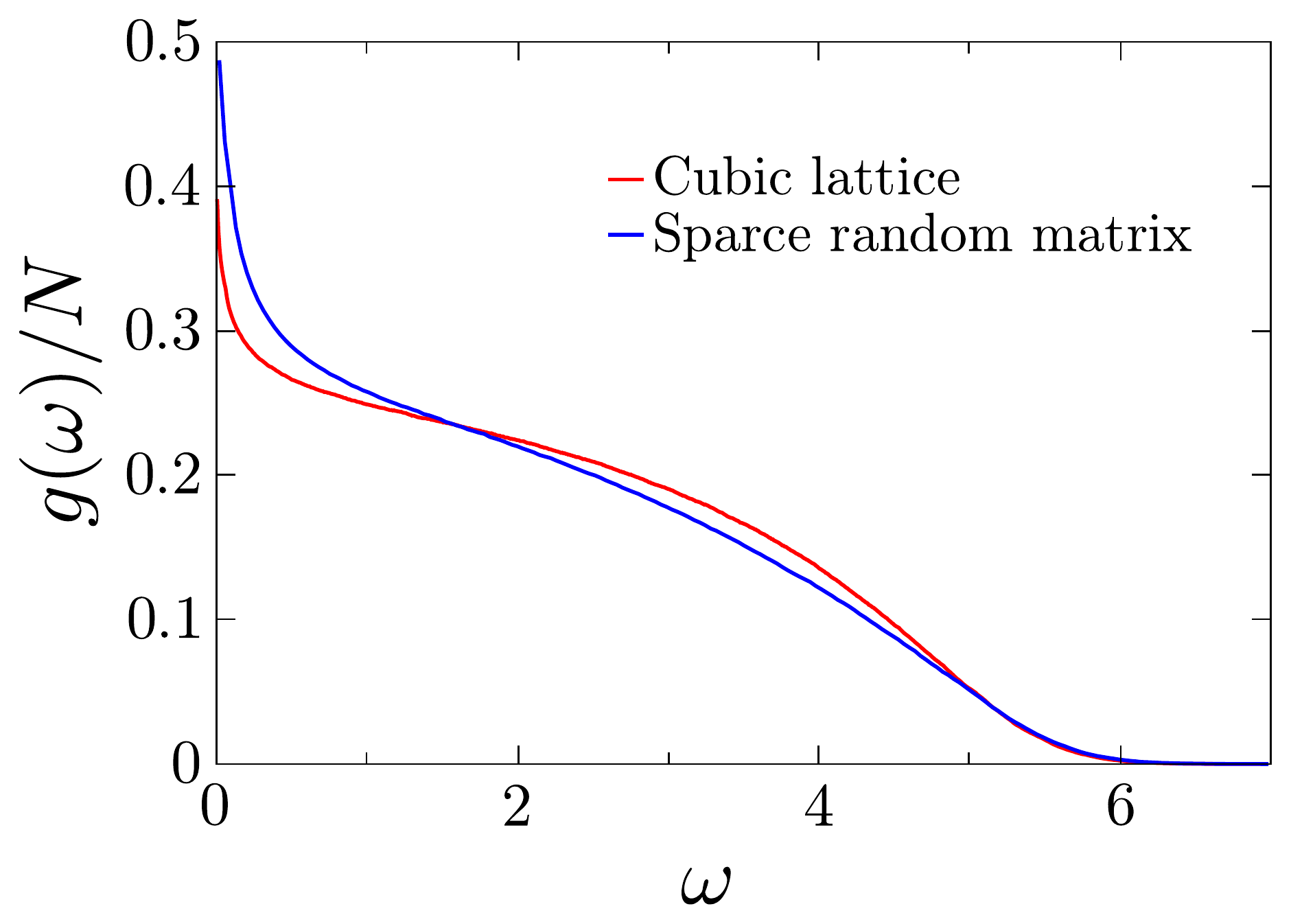}
    \caption{Comparison of the density of states for the model of a simple cubic lattice with random bonds with the model of sparse random matrix $A$ with parameter $n = 6$. In both cases $N = 1000$.}
     \label{fig:QVSS}
\end{figure}

We calculated numerically the density of vibrational states $g(\omega)$ for this cubic lattice. The mean value of
nonzero elements of the matrix $A$ was taken equal to zero, $\left<A_{ij}\right>=0$ and the variance was taken to be equal unity, $\left<A_{ij}^2\right>=1$ (Gaussian distribution). The results of the numerical calculations are
shown in Fig.~\ref{fig:QVSS}. For comparison, the results for the density of states $g(\omega)$ of the sparse random matrix $M = AA^T$ with $n = 6$ are also presented in Fig.~\ref{fig:QVSS}.
In can be seen from the figure that the density of states for the cubic lattice with random
bonds almost coincides with the density of states for the sparse random matrix with the corresponding
value of $n$. It can be shown that with increase of the radius of the interaction between atoms (i.e. if we take into account the interaction with next neighbor shels) the shape of the density of states $g(\omega)$
will approach a quarter-circle form.

\section{Participation ratio}

One of the most important problems in physics of disordered systems is the problem of modes localization.
As is well known from the seminal paper of Anderson~\cite{Anderson} a sufficiently strong disorder in the system leads to localization of elementary excitations. To estimate the inverse strength of mode localization one usually introduces the {\em participation ratio}. As a rule the participation ratio is defined as
\begin{equation}
\label{cvr9}
     P(\omega)=\frac{1}{N\sum\limits_{i=1}^{N}e_i^4(\omega)},
\end{equation}
where $e_i(\omega)$ is the $i$th coordinate of the eigenvector corresponding to the eigenvalue $\omega^2$ of the dynamical matrix $M$. In the case of completely localized mode
\begin{equation}
     |e_1|=1,\quad e_2=e_3=...=e_N=0, \quad P\sim\frac{1}{N}
     \label{Localize}
\end{equation}
the participation ratio $P(\omega)$ decreases with increase of the system size. In the case of completely
delocalized mode
\begin{equation}
     |e_1|=|e_2|=...=|e_N|=\frac{1}{\sqrt{N}}, \quad P\approx1,
     \label{Delocalize}
\end{equation}
the participation ratio does not depend on the system size $N$ and is of the order of unity.

\begin{figure}[here!]
     \centering
          \includegraphics[width=0.6\textwidth]{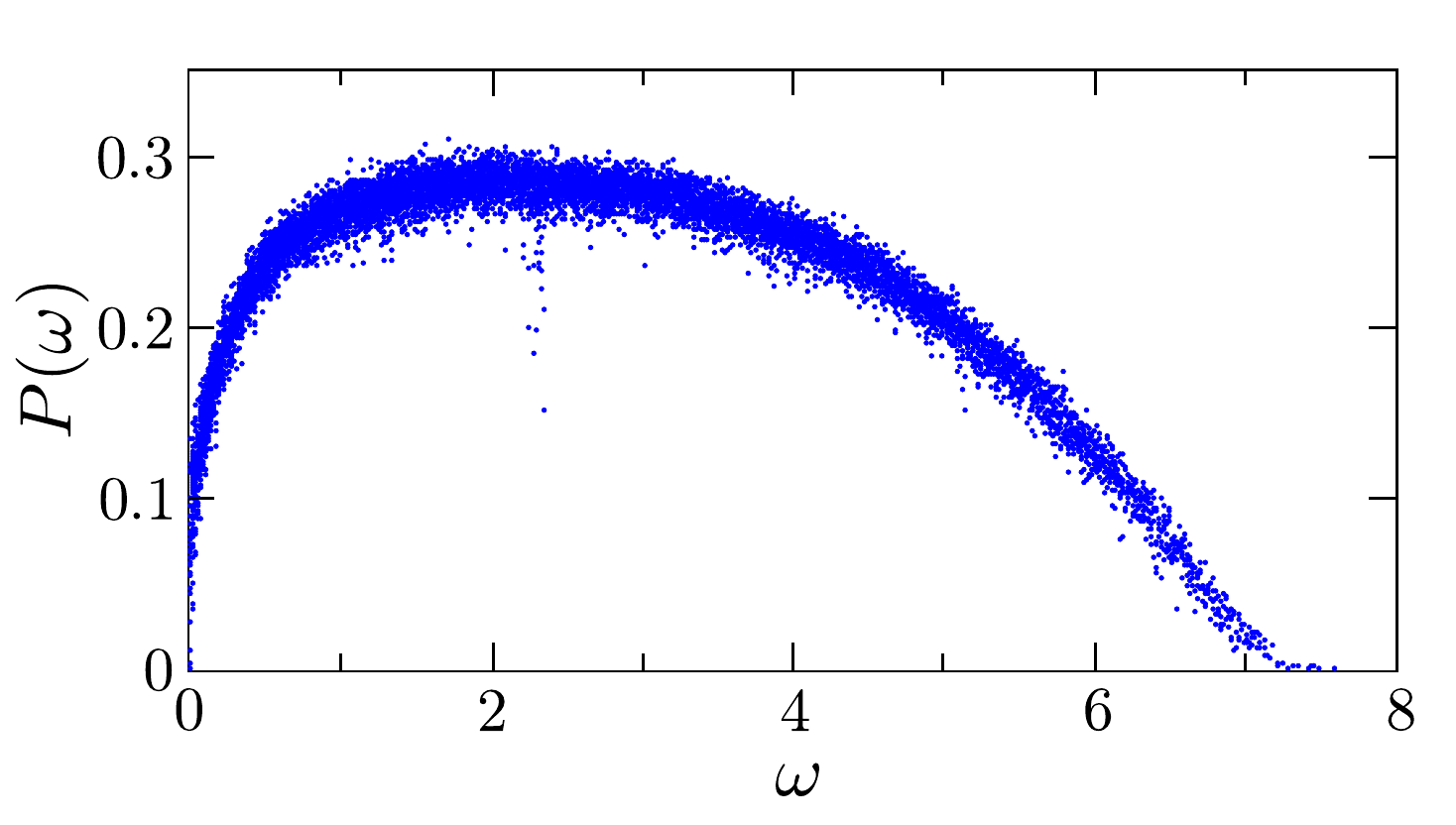}
         \caption{Numerical calculations of the participation ratio for the sparse random matrix $A$ ($M = AA^T$) for $N = 10000$ and $n = 10$. Nonzero elements of the matrix $A$ have a Gaussian distribution with
         zero mean and unit variance.}
     \label{fig:fig3}
\end{figure}
We performed numerical calculations of the participation ratio for vibrational excitations in our model.
The results for the dynamical matrix $M = AA^T$ in the case of sparse random matrix $A$
are presented in Fig.~\ref{fig:fig3}. As can be seen from this figure the participation ratio $P(\omega)$
almost for all frequencies excluding highest and lowest ones are in the range $0.2 \lesssim P(\omega) \lesssim 0.3$. One can show that it is independent of $N$. Therefore almost all vibrational modes in this range are delocalized.
Qualitatively our plot for $P(\omega)$ coincides well with the results of numerical calculations of the participation ratio for amorphous SiO$_2$ using molecular dynamics methods~\cite{Wei} in the wide frequency range $0 < \omega < 120\,$\, meV.

As it could be expected we found that an increase of parameter $n$ leads to a decrease in the number of localized modes and to an increase of the number of delocalized modes. The participation ratio for all frequencies approaches the limit equal $1/3$ (see below). Since these delocalized modes are not plane waves (phonons), they according to the terminology proposed by Allen et al.~\cite{Nature} are referred to as  {\em diffusons}.
We have shown that their spacial structure is random and character of spreading in space obeys to diffusion law. Their investigation can shed light on the mechanism of thermal conductivity in amorphous solids. This problem will be analyzed in more details in a separate work.

Similar results were obtained for the cubic lattice with random bonds. In this case the participation
ratio $P(\omega)\approx 0.2$ and is slightly lower than that in the previous case. But it does not depend on the system size $N$ as well, what indicates to delocalization of the modes.
It is interesting to note that in a two-dimensional (square) and one-dimensional lattices with random bonds (constructed
in a similar way) the participation ratio is one order of magnitude smaller than that in the 3d cubic lattice.
By analogy with disordered electronic systems~\cite{Abrahams} this can indicate to the localization of vibrational modes in these low-dimensional structures.

The numerical values of the participation ratio for vibrational modes $P(\omega)$ in various glasses according to
the data obtained by molecular dynamics methods usually are in the range $0.2\lesssim P(\omega)\lesssim 0.6$~\cite{Wei,Schober1,Schober93,Hafner,Meshkov,Ballone,Abraham}. This is in a good agreement with the results of  random matrix theory.
Indeed let us assume that the eigenvectors $e_i(\omega)$ ($i=1,2,...,N$) of the random
matrix $M = AA^T$ (which are unit vectors in $N$-dimensional space)
\begin{equation}
\label{x7yy}
\sum\limits_{i=1}^N e_i^2(\omega)=1,
\end{equation}
are isotropically distributed in all possible directions. Then the quantity $r=e_i^2(\omega)$ will be distributed
according to the Porter-Thomas law~\cite{Haake}
\begin{equation}
\label{x7xx}
p(r)=\sqrt{\frac{N}{2\pi r}}\exp\left(-\frac{Nr}{2} \right) .
\end{equation}
As a result we have
\begin{equation}
\label{x7xc}
\left<e_i^2(\omega) \right> = \left< r\right> = \frac{1}{N},\quad
\left<e_i^4(\omega) \right> = \left< r^2\right> = \frac{3}{N^2}
\end{equation}
and according to Eq.~(\ref{cvr9}) the participation ratio is
\begin{equation}
\label{x7zy}
P(\omega) = 1/3.
\end{equation}

In the literature~\cite{Schober1} one can find another definition of the participation ratio for the vector model (in contrast to the above scalar model)
\begin{equation}
     \label{Pw3}
     P_3(\omega) = \frac{1}{N \sum\limits_{i=1}^{N} \left( \sum\limits_{\alpha=1}^{3}e_{i\alpha}^2(\omega)\right)^2},
\end{equation}
where index $i$ indicates the index number of the atom ($i=1,2,...,N$) and the index $\alpha$ stands for the Cartesian projection of the displacement of the $i$th atom onto the $\alpha$ axis ($\alpha=x,y,z$).
In this case under the same assumption that the unit vectors $e_{i\alpha}(\omega)$ are isotropically distributed in $3N$-dimensional space we have in analogy with Eqs.~(\ref{x7xc})
\begin{equation}
\label{x7yc}
\left<e_{i\alpha}^2(\omega) \right> = \left< r\right> = \frac{1}{3N},\quad
\left<e_{i\alpha}^4(\omega) \right> = \left< r^2\right> = \frac{3}{(3N)^2}=\frac{1}{3N^2} .
\end{equation}
Since
\begin{equation}
\label{z7yc}
\left<\left(e_{ix}^2+e_{iy}^2+e_{iz}^2 \right)^2 \right> = \left< e_{ix}^4\right> + \left<e_{iy}^4\right> + \left<e_{iz}^4\right> +
2\left(\left<e_{ix}^2\right> \left<e_{iy}^2\right> +  \left<e_{ix}^2\right> \left<e_{iz}^2\right> + \left<e_{iy}^2\right> \left<e_{iz}^2\right>\right) ,
\end{equation}
then using Eqs.~(\ref{x7yc}) we find that the participation ratio $P_3(\omega)$ (\ref{Pw3}) is equal
\begin{equation}
     \label{cv33}
     P_3(\omega) = 3/5 = 0.6 .
\end{equation}
In this sense the values of $P(\omega)=1/3$ for the scalar model and $P_3(\omega)=0.6$ for the vector model are equivalent to each other from the standpoint of the random matrix theory. The value of $P_3(\omega)\approx 0.6$ was obtained in numerical calculations of the participation ratio for a soft-sphere glass~\cite{Laird}. Finally making use of Eqs.~(\ref{x7yc}) it can be shown that participation ratios
$P_{\rm O}\approx P_{\rm Si}\approx 0.3$ calculated numerically by Jin et al.~\cite{Wei} for amorphous $\rm{SiO_2}$ are also in a good agreement with theoretical values $P_{\rm O} = P_{\rm Si}=1/3$ that follows from the formula (18) of this work. Summarizing we can conclude that the participation ratio calculated in different papers for different glasses are in a good agreement with predictions of the random matrix theory.

\section{Level statistics}

The level statistics is another powerful criterion that makes it possible to judge about the localization or the delocalization of vibrational modes. If modes are localized their frequencies are randomly distributed over the frequency axis without any correlation with each other according to the Poisson distribution.
For quantitative description let us introduce the normalized difference between eigenfrequencies
\begin{equation}
     s = \frac{\Delta \omega}{\langle \Delta \omega\rangle}.
\end{equation}
Here $\Delta\omega=\omega_{i+1}-\omega_i$ is the distance between two neighboring frequencies that corresponds to the frequency $\omega\approx\omega_i$ and $\langle \Delta \omega\rangle$ is the mean distance between these frequencies. Then for localized modes the distribution function $Z(s)$ can be presented in the form
\begin{equation}
Z(s)=\exp(-s)
\label{ghe5}
\end{equation}
and described by the Poisson distribution.

When modes are delocalized the term repulsion effect takes place and for small values of $s\ll 1$ we
have $Z(s)\propto s$. In the random matrix theory for the Gaussian orthogonal ensemble the level statistics is described well by the Wigner surmise~\cite{Mehta,Tulino}
\begin{equation}
     Z_W(s) = \frac{\pi}{2}\,s\,\exp\left(-\frac{\pi}{4}s^2\right).
     \label{df66}
\end{equation}
As was shown by Plerou et al.~\cite{Plerov} this formula also adequately describes the Wishart ensemble.

Now let us examine analytically the problem of vibrational terms repulsion in our case. We want to find out how the probability of finding two close neighboring eigenfrequencies depends on the distance between them if this distance is much smaller than the average distance (i.e. $s\ll 1$).
For this purpose we derive the probability of finding two close neighboring eigenvalues
$\varepsilon_i$ and $\varepsilon_j$ of the matrix $M = AA^T$ that are the squares of the corresponding eigenfrequencies. We will restrict our analysis to the case of Wishart ensemble considered in Section~\ref{par:quarter}.

In  Eq.~(\ref{Perturbation4}) the term in the sum that contains the difference between these two close
eigenvalues in the denominator will be considerably larger than the other terms. Let us take into account only this term. Then
\begin{equation}
     \delta\varepsilon\left(\varepsilon_i, V^2\right) = \frac{2v^2\varepsilon_i}{\varepsilon_i-\varepsilon_j}.
     \label{Perturbation4easy}
\end{equation}
Let $\nu = \varepsilon_i-\varepsilon_j>0$ be the difference between two neighboring eigenvalues.
Then its change due to the perturbation $\delta A$ of the matrix $A$ is given by
\begin{equation}
     \delta\nu\left(\nu, \varepsilon, V^2\right) = \delta\varepsilon\left(\varepsilon_i, V^2\right) - \delta\varepsilon\left(\varepsilon_j, V^2\right)= \frac{4v^2\varepsilon}{\nu},
     \label{PerturbationS}
\end{equation}
where $\varepsilon$ is the mean value between  eigenvalues $\varepsilon_i$ and $\varepsilon_j$.

Let $z(\nu, \varepsilon, V^2)$ be the density of distribution of the difference $\nu$ between two neighboring eigenvalues. Similar to the derivation of Eq.~(\ref{incinc}) we get
\begin{equation}
     -\frac{\partial(z \delta \nu)}{\partial\nu} = v^2\frac{\partial z}{\partial (V^2)}.
     \label{diffEq2}
\end{equation}
The function $z(\nu, \varepsilon, V^2)$ can be presented in a form
\begin{equation}
     z\left(\nu, \varepsilon, V^2\right) = \frac{1}{V^2}\widetilde{z}(x, \varepsilon),
     \label{tildeZ}
\end{equation}
where $x = \nu/V^2$. Substitution of Eqs.~(\ref{PerturbationS}, \ref{tildeZ}) into Eq.~(\ref{diffEq2}) gives
\begin{equation}
     \frac{4 \varepsilon}{V^2}\left(x\frac{\partial\widetilde{z}}{\partial x}-\widetilde{z}\right) = x^2\left(\widetilde{z}+x\frac{\partial\widetilde{z}}{\partial x}\right).
\end{equation}
For small values of $x$ the right-hand side of this equation can be replaced by zero. In this case its solution
for the function $\widetilde{z}(x)$ will be proportional to $x$.
This in turn means that for small $\nu$ the function $z(\nu, \varepsilon, V^2)\propto \nu$ i.e. there is a term repulsion effect.

\begin{figure}[htbp]
     \centering
     \includegraphics[width=0.6\textwidth]{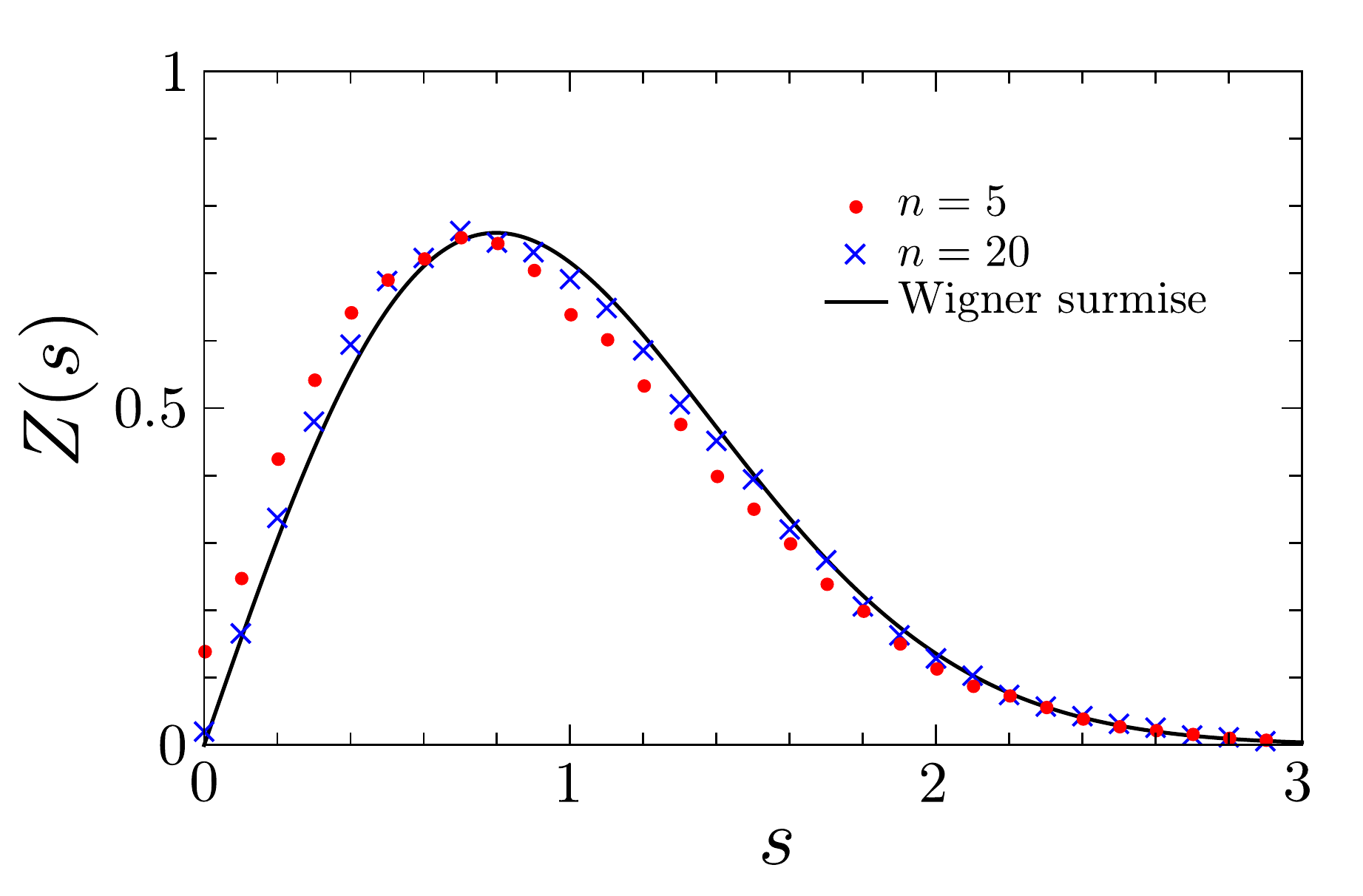}
     \caption{Numerical calculation of the density of distribution of the difference between neighboring frequencies i.e. square roots of the eigenvalues of the matrix $M = AA^T$. The matrix $A$ is a sparse random matrix with
     $N = 1000$ and the number of nonzero elements $n$ in each row. Nonzero elements of the matrix $A$ have a Gaussian distribution with zero mean and unit variance. The solid line indicates the theoretical Wigner dependence (\ref{df66}).}
     \label{fig:fig2}
\end{figure}

In the case of the sparse random matrix $A$ we have calculated numerically the level statistics for the dynamical matrix $M = AA^T$. The result (averaged over all frequencies) is shown in Fig.~\ref{fig:fig2}.
It follows from this figure that for small $s\ll 1$ we indeed have $Z(s) \propto s$ what is in agreement with the analytical result. It can also be seen from the Fig.~\ref{fig:fig2} that Wigner formula (\ref{df66}) adequately describes the level statistics in the case of sparse matrices. As was mentioned above such statistics corresponds to the case of delocalized modes. Therefore we conclude that majority of the vibrational modes in our system are delocalized. This is in a good agreement with the data presented in the Fig.~\ref{fig:fig3} for the participation ratio $P(\omega)$. In conclusion we note that our results agree well with the molecular dynamics calculations of the statistics of vibrational levels for amorphous clusters~\cite{Sarkar1,Sarkar2}.

\section{The distribution of elements of the dynamical matrix}

In the last section we consider the statistics of nonzero matrix elements of the dynamical sparse random matrix $M=AA^T$ and compare it with available computer calculations performed for amorphous systems by molecular dynamics methods. First we analyze the distribution of nonzero off-diagonal elements of the matrix $M$. As follows from the Eq.~(\ref{vst5rf}) an off-diagonal matrix element $M_{ij}$ is the ``scalar product'' of the $i$th and $j$th rows of the matrix $A$. In order for the element $M_{ij}$ at $i\ne j$ to be nonzero it is necessary that the column positions of at least one pair of nonzero elements  in the $i$th and $j$th rows of the matrix $A$ coincide. For a sparse random matrix $A$ (i.e. for $n\ll N$) the probability of coincidence of more than one pair of nonzero elements is
negligible. Therefore the off-diagonal elements of the matrix $M$ for the most part are equal to zero and if
they are nonzero they are the products of two independent nonzero elements of the matrix $A$.
Then the density of distribution of nonzero off-diagonal elements of the matrix $M$ is given by
\begin{equation}\label{RhoMult}
     \rho_M^{\rm off}(x) =
     \int_{-\infty}^\infty\int_{-\infty}^\infty\delta(x-\xi\eta)\rho(\xi)\rho(\eta)d\xi d\eta
     =\int_{-\infty}^\infty\frac{1}{|t|}\rho\left(\frac{x}{t}\right)\rho(t)dt.
\end{equation}
Here $\rho(x)$ is the distribution function of nonzero elements of the random matrix $A$.
In the case when the non-zero elements of the matrix $A$ are normally distributed with zero mean and variance $V^2$ we have
\begin{equation}\label{RhoNorm}
     \rho(x)=\frac{1}{V\sqrt{2\pi}}\exp\left(-\frac{x^2}{2V^2}\right).
\end{equation}
Substitution of the probability density (\ref{RhoNorm}) into the integral (\ref{RhoMult}) gives the integral representation of the zeroth order Macdonald function
\begin{equation}
\label{hv6}
     \rho_M^{\rm off}(x) =\frac{1}{\pi V}K_0(|x|/V).
\end{equation}

\begin{figure}[htbp]
     \centering
     \includegraphics[width=0.60\textwidth]{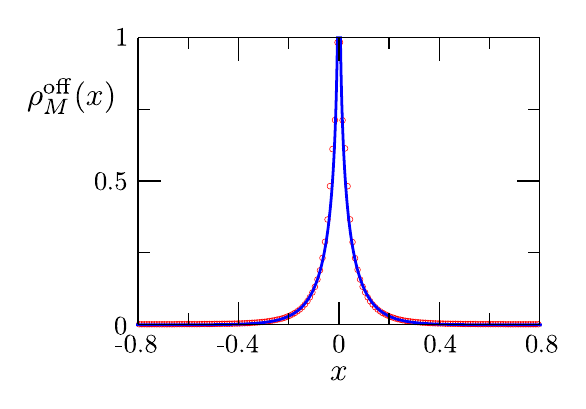}
     \caption{Comparison of the distribution (\ref{hv6}) (solid line) with the results of the numerical calculations~\cite{Huang} (open circles). The variance is $V^2=0.01$.}
     \label{fig:K0}
\end{figure}
We have compared this function with numerical results of the paper~\cite{Huang} where the dynamical matrix for a disordered system of atoms interacting through the Lennard-Jones potential was calculated using the molecular dynamic methods. It can be seen from the Fig.~\ref{fig:K0} that for the variance $V^2 = 0.01$ our Eq.~(\ref{hv6}) is in a good agreement with the numerical results~\cite{Huang}.

\begin{figure}[!here]
     \centering
     \includegraphics[width=0.60\textwidth]{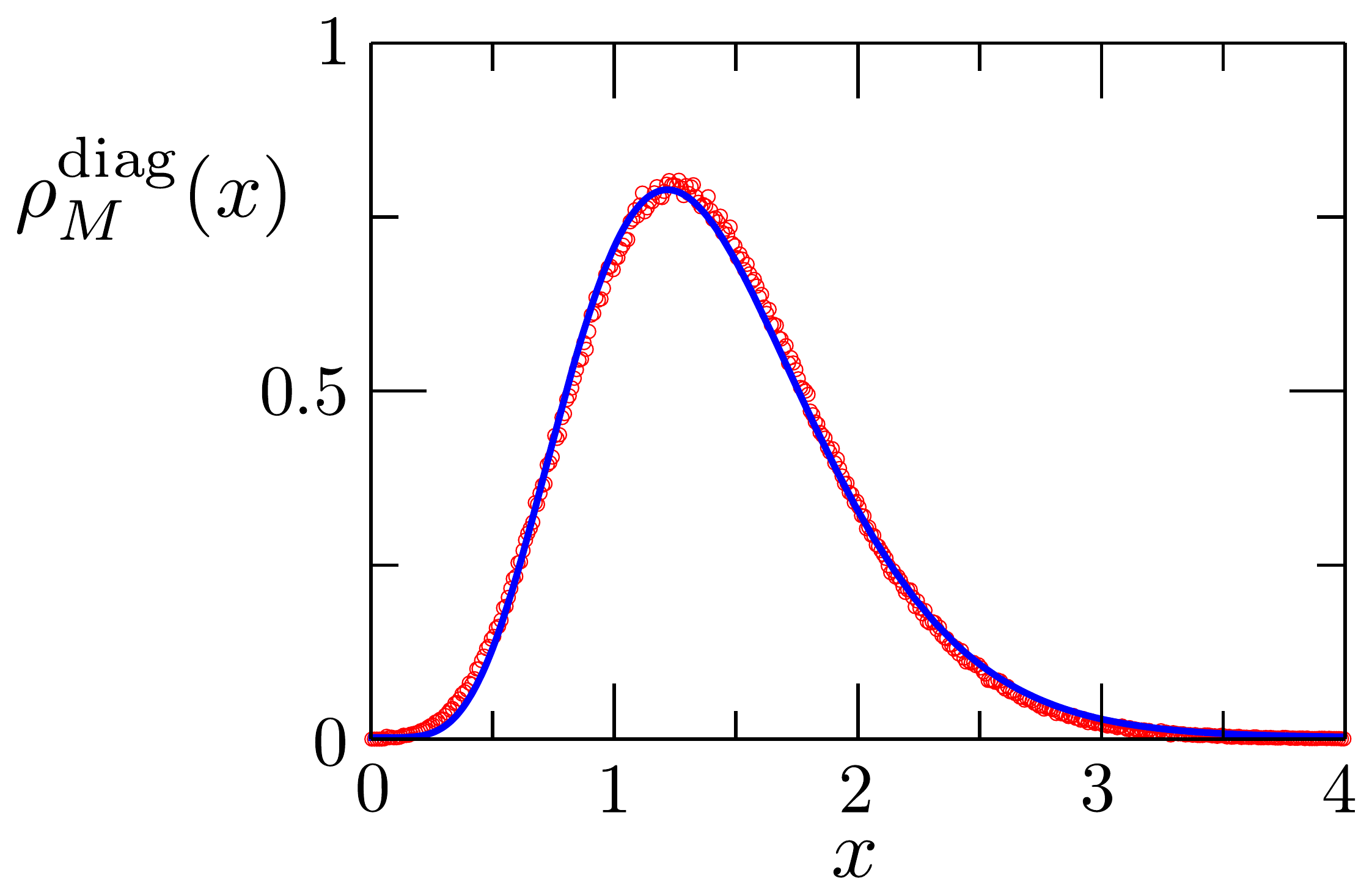}
      \caption{Comparison of the distribution (\ref{Chi}) of the diagonal elements of the matrix $M = AA^T$ for $n = 14$ and $V^2 = 0.01$ (solid line) with the results of the numerical calculations~\cite{Huang} (open circles).}
     \label{fig:Chi}
\end{figure}

The diagonal elements of the matrix $M$ are distributed as a sum of $n$ squares of elements from one row of the matrix $A$. When the elements of the matrix $A$ are normally distributed with zero mean and variance $V^2$ we have well known $\chi^2$ distribution of diagonal elements of the matrix $M$
\begin{equation}
     \rho_M^{\rm diag}(x) = \frac{(1/2)^{n/2}}
     {V\Gamma \left(n/2\right)}(x/V)^{n/2-1}\exp\left(-\frac{x}{2V}\right).
     \label{Chi}
\end{equation}
As it can be seen from the Fig.~\ref{fig:Chi} this distribution (for the same choice of the variance $V^2 = 0.01$ and $n=14$) also agrees quite well with the results of the numerical calculations in~\cite{Huang}.

\section{Conclusion}

We have demonstrated that the dynamical sparse random matrices of the general form
$M = AA^T$ with nonnegative eigenvalues $\varepsilon=\omega^2$ can be successfully used for the description of sufficiently general properties of the vibrational spectra of amorphous solids. Compared to the currently used molecular dynamics methods they have an important advantage that the construction of the random dynamical matrix corresponding to a stable system requires significantly less efforts than numerical calculations for real glasses with specific interatomic interaction potentials. In many cases the results are found to be quite similar.

The main our result is that the density of states $g(\omega)$ of the vibrational system described by the
sparse random matrix $M = AA^T$ is given by the quarter-circle law. In our model this form of the density of vibrational states is a universal law. It depends neither on the density of distribution of non-zero matrix elements of the sparse random matrix $A$ (provided that their mean is equal to zero and the variance is finite), nor on the size of the system $N$, nor on the number of nonzero elements $n$ in each row of matrix $A$ for $n\gg 1$. This result also does not depend strongly on the topology of elastic bonds between the atoms. It was shown for disordered systems with topology of a tree (Bethe lattice) and for a cubic lattice with random bonds.

The study of the problem of localization of these vibrational modes in the three-dimensional system led us to the conclusion that despite a high degree of disorder the majority of the modes are delocalized harmonic excitations.
This is evidenced by the values of the participation ratio and the statistics of vibrational levels where the term repulsion effect clearly manifests itself. Our results are in a good agreement with the results obtained for real glasses by molecular dynamics methods.

Finally by using the concept of sparse random matrices we calculated the distribution of matrix elements of the dynamical matrix $M$. The results of the calculations are  in a good agreement with the numerical data obtained by molecular dynamics methods for some class of amorphous systems. The aforesaid allows us to draw the important conclusion that random dynamical matrices of the type $M = AA^T$ reflect some universal characteristics of vibrational spectra of amorphous solids.

In conclusion it should be noted that the density of states in the form of the quarter-circle distribution
obtained in our work means that $g(\omega)$ tends to a constant value at $\omega\to 0$. However usually in disordered systems (in three-dimensional space) such behavior of $g(\omega)$ can occur only in the absence of the low frequency acoustical phonons. These phonons are weakly decayed plane wave excitations propagating in space with velocity of sound. Therefore as a rule  the density of states of glasses at low frequencies has a {\em phonon gap}, $g(\omega)\propto \omega^2\to 0$ at $\omega\to 0$. As we already mentioned  these phonon modes usually contain not more than 10\% of all vibrational modes. At higher frequencies we have delocalized vibrations of different type, namely  diffusons. They occupy the larger part of the spectrum. We believe that we can use random matrix theory to describe them.

Relatively recently in the literature there have been appeared disordered systems of a new type.
In these systems the width of the phonon gap can be decreased or even reduced to zero by decreasing the
rigidity of the system. In the latter case it was found that the density of vibrational modes $g(\omega)$ indeed
goes to a constant value at $\omega\to 0$. Such behavior of $g(\omega)$ was observed in computer simulations of disordered systems of the type of granular media in the vicinity of the {\em jamming transition} point when decreasing the density of particles interacting through repulsive forces with a finite effective radius~\cite{Nagel,Hecke}.
A similar behavior was also observed for other computer models~\cite{Ashton}. Let us mention also the calculations
performed by Trachenko et al.~\cite{Trachenko} who calculated the density of vibrational states in amorphous $\rm{SiO_2}$ in the rigid tetrahedron approximation. In this approximation the density of states $g(\omega)$ also was found to be finite at $\omega\to 0$. We hope that our random matrix model also will make it possible to describe such systems.

\section{Acknowledgements}

We would like to thank V.L. Gurevich and V.I. Kozub for numerous helpful discussions.
One of authors (Y.B.) thanks the St. Petersburg Government for the financial support (diploma project no. 2.4/04-05/103).

\end{document}